\documentclass[prc,twocolumn,showpacs]{revtex4}
\usepackage{graphicx}
\usepackage{dcolumn}
\usepackage{bm}

\begin{document}

\title{Contribution of excited states to stellar weak-interaction rates in odd-A nuclei}

\author{P. Sarriguren}
\affiliation{Instituto de Estructura de la Materia, IEM-CSIC, 
Serrano 123, E-28006 Madrid, Spain}

\email{p.sarriguren@csic.es}

\date{\today}

\begin{abstract}
 
Weak-interaction rates, including $\beta$-decay and electron capture, are studied 
in several odd-$A$ nuclei in the {\it pf}-shell region at various densities and 
temperatures of astrophysical interest. Special attention is paid to the relative 
contribution to these rates of thermally populated excited states in the decaying 
nucleus. The nuclear structure involved in the weak processes is studied within a 
quasiparticle random-phase approximation with residual interactions in both 
particle-hole and particle-particle channels on top of a deformed Skyrme 
Hartree-Fock mean field with pairing correlations. In the range of densities and 
temperatures considered, it is found that the total rates do not differ much from 
the rates of the ground state fully populated. In any case, the changes are not 
larger than the uncertainties due to the nuclear model dependence of the rates.

\end{abstract}


\maketitle

\section{Introduction}

The relevant role that weak $\beta$-decay and electron-capture (EC) processes
play in our understanding of the late stages of the stellar evolution has been 
long recognized \cite{b2fh,ffn}. In particular, the presupernova stellar 
structure, as well as the nucleosynthesis of heavier nuclei, are determined 
to a large extent by those mechanisms. {\it pf}-shell nuclei are specially 
important in these scenarios because they are the main constituents of the 
stellar core in presupernova formations \cite{aufderheide-94} leading to 
core-collapse (type II) or thermonuclear (type Ia) supernovae.

Type II supernovae are thought to be the final result of the gravitational 
collapse of the core of a massive star that takes place when the nuclear fuel 
exhausts. In the initial stages of the collapse, electrons are captured by 
nuclei in the iron-nickel mass region, thus reducing the electron-to-baryon 
fraction ($Y_e$) of the presupernova star and correspondingly the degeneracy 
pressure. At the same time, at typical presupernova densities, the neutrinos 
generated in those captures can leave the star, reducing the energy and cooling 
the star. Both effects help to accelerate the stellar collapse. EC processes 
are therefore essential ingredients to follow the complex dynamics of 
core-collapse supernova, and reliable estimates of these rates certainly 
contribute to a better understanding of the explosion mechanism.

Network calculations and astrophysical simulations \cite{langanke-03} 
require nuclear physics input that in most cases cannot be measured directly
in the laboratory due to the extreme conditions of densities ($\rho$) and 
temperatures (T) holding in stellar scenarios. Therefore, nuclear properties, 
and in particular the Gamow-Teller (GT) transitions that determine to a large 
extent the decay properties, must be estimated in many cases by model 
calculations.

A strong effort has been made in the last decades to measure the GT strength 
distributions of nuclei in the mass region $A \sim 60$. This has been performed
by means of $(n,p)$ or equivalent higher resolution charge-exchange reactions 
such as $(d,^2He)$ and $(t,^3He)$ at forward angles \cite{fujita-11}. These 
reactions are the most efficient way to extract the GT strength in stable 
nuclei \cite{osterfeld-92}.

From the theoretical side, the first extensive calculations of stellar weak 
rates in relevant ranges of $\rho$ and T were done in Ref. \cite{ffn} under 
severe assumptions concerning the energy distribution of the GT strength.
In Ref. \cite{ffn} the whole GT strength was assumed to be concentrated in a 
single resonance at an energy parametrized phenomenologically. The total GT 
strength was taken from the single-particle model. Since then, improvements in 
the weak rates have been focused on the description of the nuclear structure 
aspect of the problem. Different approaches are found in the literature that 
can be roughly divided into shell model (SM) 
\cite{koonin-97,dean-98,langanke-00,suzuki-09} and proton-neutron quasiparticle 
random-phase approximation (QRPA) \cite{krumlinde-84,moeller-95,moeller-97,
muto-89,nabi-99,nabi-09,paar-04,paar-09,fantina-12,dzhioev-10,niu-11,sarri-98,
sarri-99,sarri-01-odd,sarri-13} categories. Hybrid models using RPA methods 
on top of a temperature-dependent description of the parent nucleus using a 
shell-model Monte Carlo approach have been also developed to calculate stellar 
EC rates \cite{langanke-01}.
Although QRPA calculations cannot reach the detailed spectroscopic 
accuracy achieved by present state-of-the-art SM calculations, the global 
performance of QRPA is quite satisfactory. Moreover, one clear advantage of
the QRPA method is that it can be extended to heavier nuclei, which are beyond
the present capability of full SM calculations, without increasing the 
complexity of the calculation. 

A systematic evaluation of the ability to 
reproduce the measured GT strength distributions of various theoretical models 
based on SM and QRPA was done in Ref. \cite{cole-12}. EC rates were also derived 
from those models at astrophysical conditions of $\rho$ and T. In Ref. 
\cite{cole-12} SM calculations using different effective interactions were 
compared with data and with QRPA of Ref. \cite{moeller-97}. Later in Ref. 
\cite{sarri-13} the same results were also compared with QRPA calculations 
using a Skyrme selfconsistent mean field and QRPA with residual interactions 
in both particle-hole (ph) and particle-particle (pp) channels, improving 
significantly the agreement with experiment.

It is important to realize that there are clear differences between terrestrial 
and stellar decay rates caused by the effect of high $\rho$ and T conditions. 
One effect of T is directly related to the thermal population of excited states 
in the decaying nucleus, accompanied by the corresponding depopulation of the 
ground states. The weak rates of excited states can be significantly different 
from those of the ground state and a case by case consideration is needed.
Another distinctive effect comes from the fact that atoms in stellar scenarios 
are completely ionized, and consequently electrons are no longer bound to the 
nuclei, but forming a degenerate plasma that obeys a Fermi-Dirac distribution.
This opens the possibility for continuum EC, in contrast to the orbital EC 
caused by bound electrons in the atom under terrestrial conditions. These two 
effects make weak-interaction rates in the stellar interior sensitive functions 
of T and $\rho$.

In addition to these genuine stellar effects, one has to deal with uncertainties 
in the extraction of the experimental GT strength due to several causes such as 
the global normalization of the unit cross section, or to possible interference 
effects caused by the tensor component of the interaction. One has to take also 
into account that the GT strength is only measured up to some excitation energy 
and therefore the rates calculated from it will not include possible 
contributions from transitions beyond the measured energy range that could have 
an effect, especially at high T and $\rho$. The measured GT strength distributions 
are then strict tests to constrain the theoretical ones under terrestrial 
conditions, but the  rates calculated from the experimental GT strength 
distributions may still differ significantly from the actual rates operating 
in stars.

The nuclei under study in this work correspond in most cases to stable 
{\it pf}-shell nuclei and $\beta^+$ decays from their ground states are 
energetically forbidden. However, as T raises, thermal population of excited 
states in the parent nucleus may induce $\beta ^+$ decays if the energy 
excitation in the parent nucleus exceeds the $Q_\beta$ energy. These decays, 
which are almost independent of $\rho$ and T, might compete with ECs in 
particular cases. Competition of ECs with  $\beta ^-$ decays in somewhat 
neutron-rich nuclei in this mass region have also been studied in 
Ref. \cite{pinedo-00}.

In the case of even-even nuclei studied in Ref. \cite{sarri-13} the eventual 
contributions from excited states could be safely neglected because of the 
high energy excitation of the first excited states, typically $2^+$ states 
beyond 1 MeV that can hardly be excited within the range of temperatures 
considered in this work. However, low-lying excited states would contribute
to the rates in the case of even-even well deformed nuclei \cite{sarri-09-11}, 
where the rotational states drop easily below 1 MeV, as well as in odd-$A$ 
nuclei where quasiparticle states are found at very low excitation energies. 
The main purpose of this work is to study these contributions to the weak 
rates coming from low-lying excited states in odd-$A$ nuclei. 

To perform this study, a set 
of nuclei has been chosen according to their interest in presupernova models 
\cite{aufderheide-94,cole-12}. This set of nuclei includes $^{45}$Sc, $^{51}$V, 
$^{53}$Fe, $^{55}$Fe, $^{55}$Mn, and $^{57}$Fe. They exhibit 
in most cases rich low-lying spectra with several excited states below 1 MeV 
that are displayed in Fig. \ref{en_exp}. There is also experimental information 
\cite{alford-91,alford-93,baumer-03,elkateb-94} on the GT strength distributions
in $^{45}$Sc, $^{51}$V, and $^{55}$Mn obtained from charge-exchange 
reactions that will be used for comparison. 

The structure of the paper is as follows. After presenting briefly the 
theoretical framework used to study the weak-interaction rates and the nuclear 
structure involved in Sec. II, a comparison of the results with the available 
measurements of the energy distribution of the GT strength will be performed 
in Sec. III.A. The weak rates will be evaluated in Sec. III.B at various stellar 
conditions including explicitly the contributions of the excited states. The 
main conclusions of the work are presented in Sec. IV.

\section{Theoretical Formalism}
\label{form}

\subsection{ Weak-interaction rates}  
\label{wdr}

The weak-interaction rate of a nucleus can be expressed as follows,

\begin{equation}
\lambda = \sum_i \lambda_i P_i(T);\quad P_i(T)=\frac{2J_i+1}{G} e^{-E_i/(k_BT)} \, ,
\label{population}
\end{equation}
where $P_i(T)$ is the probability of occupation of the excited state $i$ in the 
parent nucleus. Assuming thermal equilibrium, it is given by a Boltzmann 
distribution. $G=\sum_i \left( 2J_i+1 \right) e^{-E_i/(k_BT)}$ is the partition 
function and $J_i(E_i)$ is the angular momentum (excitation energy) of the parent 
nucleus state $i$. In principle, the sum extends over all excited states in 
the parent nucleus up to the proton separation energy. However, because of the 
range of temperatures considered in this work ($T=1-10$ GK), only a few low-lying
excited states are expected to contribute significantly. The scale of excitation 
energies to consider is determined by $k_BT$, which for maximum T around 10 GK 
is given by $k_BT = 0.862$ MeV. Thus, excitation energies beyond 1 MeV are not 
considered in this work. The experimental energies of the states considered can 
be seen in Fig. \ref{en_exp}.
The weak-interaction rate corresponding to the parent state $i$ is given by

\begin{equation}
\lambda _i = \sum_f \lambda_{if} =  \frac{\ln 2}{D}  \sum_f 
B_{if}\Phi_{if} (\rho,T)\, ,
\label{rate_i}
\end{equation}
where the sum extends over all the states in the final nucleus that can be 
reached in the process and $D=6146$ s. This expression is decomposed into a 
phase space factor $\Phi_{if}$, which is a function of $\rho$ and $T$ and a 
nuclear structure part $B_{if}$ that contains the transition probabilities for 
allowed Fermi and GT transitions. In this work we only consider the dominant 
GT transitions. Fermi transitions are only important for $\beta^+$ decay of 
neutron-deficient light nuclei with  $Z> N$. The theoretical description of 
both $\Phi_{if}$ and $B_{if}$ are explained in the next subsections.

Dealing with the full sums involved in Eqs. (\ref{population}) and (\ref{rate_i})
is in general impracticable, but because our goal is to evaluate the 
contributions of a few low-lying excited states in odd-$A$ nuclei to the 
total rates rather than calculating the full rates as a function of the 
temperature, a state by state calculation of the sums is feasible.
Other alternatives based on equilibrium statistical formulations of the nuclear 
many-body problem have been explored giving rise to a temperature-dependent GT 
strength function.
Thus, in Refs. \cite{paar-09,fantina-12} a fully selfconsistent microscopic 
framework was used for the calculation of weak-interaction rates at finite 
temperature, based on spherical mean-field models with Skyrme functionals and a 
finite-temperature RPA. EC were calculated from various Skyrme interactions to 
estimate the resulting theoretical uncertainty. Differences were found in the 
EC rates that can be sizable at low temperatures. In Ref. \cite{dzhioev-10}
stellar weak decay rates were studied within a QRPA approach based on a
spherical Woods-Saxon potential and separable interactions, extended to 
finite temperature by the thermofield dynamics formalism. 
A fully selfconsistent relativistic framework was also introduced in 
Ref. \cite{niu-11} to
study EC on nuclei in stellar environments. The formalism was based on a 
finite-temperature relativistic mean field with charge-exchange transitions 
described within a selfconsistent finite-temperature relativistic RPA.

\subsection{Phase Space Factors}

In the astrophysical scenarios of our study, atoms are assumed to be fully ionized and 
continuum ECs from the degenerate electron plasma are possible. The phase space 
factor for continuum EC is given by

\begin{eqnarray}
\Phi^{EC}_{if}&=&\int_{\omega_\ell}^{\infty} \omega p (Q_{if}+\omega)^2
F(Z,\omega) \nonumber \\
&& \times S_{e}(\omega) \left[ 1-S_{\nu}(Q_{if}+\omega)\right] d\omega \, .
\label{phiec}
\end{eqnarray}

We also consider the possibility of $\beta^+$ decay, not only  because some of 
the nuclei studied are $\beta^+$ unstable, as it can be seen from the positive 
values of $Q_{EC}$ in Table I, but also because excited states in
the parent nucleus can be thermally populated and thus, decay is possible for 
nuclei which are stable in terrestrial conditions. 
The phase space factor for positron emission $\beta^+$ process is given by 

\begin{eqnarray}
\Phi^{\beta^+}_{if}&=&\int _{1}^{Q_{if}} \omega p  
(Q_{if}-\omega)^2 F(-Z+1,\omega) \nonumber \\
&& \times \left[ 1-S_{e^+}(\omega)\right]
\left[ 1-S_{\nu}(Q_{if}-\omega)\right] d\omega \, .
\label{phib}
\end{eqnarray}

In these expressions $\omega$ is the total energy of the electron in $m_ec^2$ 
units, $p=\sqrt{\omega ^2 -1}$ is the momentum, and $Q_{if}$ is the total energy 
available in $m_e c^2$ units

\begin{equation}
Q_{if}=\frac{1}{m_ec^2}\left( Q_{EC} - m_e c^2 +E_i-E_f \right) \, ,
\label{qif}
\end{equation}
with 
\begin{equation}
Q_{EC} = Q_{\beta^+} + 2m_e c^2 = \left( M_p-M_d+m_e\right) c^2 \, ,
\label{qec}
\end{equation}
written in terms of the nuclear masses of parent ($M_p$) and daughter ($M_d$) 
nuclei and their excitation energies $E_i$ and $E_f$, respectively. 
$F(Z,\omega)$ is the Fermi function that takes into account the distortion of 
the electron wave function due to the Coulomb interaction.

\begin{equation}
F(Z,\omega ) = 2(1+\gamma) (2pR)^{-2(1-\gamma)} e^{\pi y}
\frac{|\Gamma (\gamma+iy)|^2}{[\Gamma (2\gamma+1)]^2}\, ,
\end{equation}
where $\gamma=\sqrt{1-(\alpha Z)^2}$ ; $y=\alpha Z\omega /p$, $\alpha$ is the 
fine structure constant, and $R$ the nuclear radius. The lower integration limit 
in Eq. (\ref{phiec}) is given by ${\omega_\ell}=1$ if $Q_{if}> -1$, or 
${\omega_\ell}=|Q_{if}|$ if $Q_{if}< -1$.

$S_e$, $S_{e^+}$, and $S_\nu$, are electron, positron, and neutrino distribution 
functions, respectively. Its presence inhibits or enhances the phase space 
available. In the stellar scenarios considered here the commonly accepted 
assumption is that $S_{e^+}=0$ because electron-positron pair creation becomes 
important only at higher energies and $S_\nu=0$ because neutrinos and 
antineutrinos can escape freely from the interior of the star at these densities. 
The electron distribution is described by a Fermi-Dirac distribution

\begin{equation}
S_{e}=\frac{1}{\exp \left[ \left(\omega -\mu_e\right)/(k_BT)\right] +1} \, .
\label{se}
\end{equation}
The chemical potential $\mu_e$ as a function of $\rho$ and $T$ is 
determined from the expression

\begin{equation}
\rho Y_e = \frac{1}{\pi^2 N_A}\left( \frac{m_e c}{\hbar}\right) ^3 
\int_0^{\infty} (S_e - S_{e^+}) p^2 dp \, ,
\end{equation}
in (mol/cm$^3$) units. $\rho$ is the baryon density (g/cm$^3$),
$Y_e$ is the electron-to-baryon ratio (mol/g), and $N_A$ is Avogadro's
number (mol$^{-1}$).

The phase space factor for EC in Eq. (\ref{phiec}) is therefore a sensitive 
function of both $\rho$ and $T$, through the electron distribution $S_e$. 
On the other hand, the phase space factor for $\beta^+$ decay in Eq. 
(\ref{phib}) under the assumptions $S_{e^+}=S_\nu=0$, does not depend on 
$\rho$ and $T$. The only dependence of the positron decay rates to $\rho$ and 
$T$ appears indirectly through the population of excited states.

\subsection{Nuclear Structure}


\begin{table}[ht]
\begin{center}
\caption{Experimental $Q_{EC}$ (MeV) values \cite{ensdf}.}
\vskip 0.5cm
\begin{tabular}{cccccc}
\hline \hline \\
  $^{45}$Sc &  $^{51}$V & $^{53}$Fe  & $^{55}$Fe & $^{55}$Mn &
  $^{57}$Fe  \\ \\
 -0.259 & -2.472 & +3.742 & +0.231 & -2.603 &  -2.659   \\ 
\hline \hline
\end{tabular}
\end{center}
\label{qectable}
\end{table}

\begin{figure}[ht]
\includegraphics[width=0.42\textwidth]{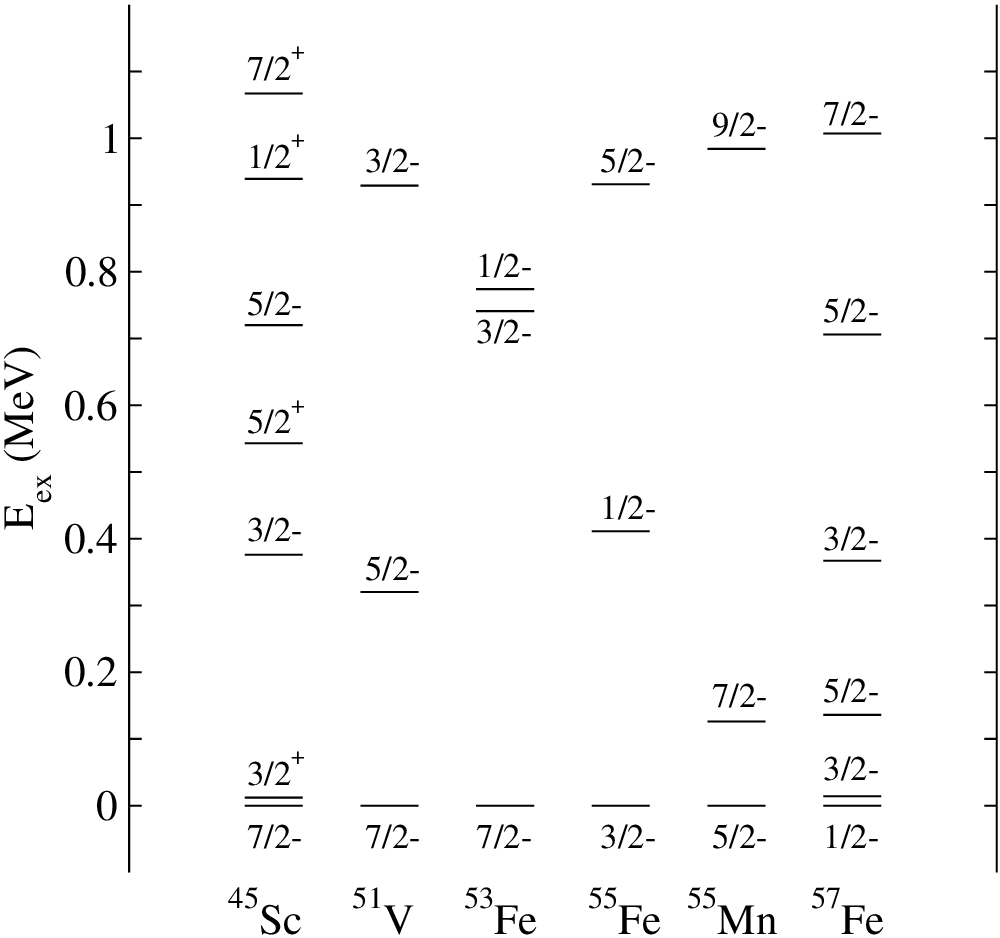}
\caption{Experimental energies of the low-lying excited states.} 
\label{en_exp}
\end{figure}

The nuclear structure part of the problem is described within the QRPA formalism. 
Various approaches have been developed in the past to describe the spin-isospin 
nuclear excitations in QRPA \cite{krumlinde-84,moeller-95,moeller-97,
muto-89,nabi-99,nabi-09,paar-04,paar-09,fantina-12,dzhioev-10,niu-11,sarri-98,
sarri-99,sarri-01-odd,sarri-13}. In this subsection 
we show briefly the theoretical framework used in this work to describe the nuclear 
part of the weak-interaction rates. More details of the formalism can be found in 
Refs. \cite{sarri-98,sarri-99,sarri-01-odd}.

The method starts with a selfconsistent deformed Hartree-Fock mean field 
calculation with Skyrme interactions including pairing correlations. 
The single-particle energies, wave functions, and occupation probabilities 
are generated from this mean field. The Skyrme force SLy4 \cite{sly4} is used 
in this work. It is one of the most successful Skyrme forces and has been 
extensively studied in the past. The sensitivity of the EC rates to different
choices of the Skyrme interactions was studied in Ref. \cite{sarri-13} within
the same formalism, using SLy4 and SGII Skyrme interactions. 
The solution of the HF equation is found by using the formalism developed in 
Ref. \cite{vautherin}, assuming time reversal and axial  symmetry. The 
single-particle wave functions are expanded in terms of the eigenstates of an 
axially symmetric harmonic oscillator in cylindrical coordinates, using twelve 
major shells. The method also includes pairing between like nucleons in BCS 
approximation with fixed gap parameters for protons and neutrons, which are 
determined phenomenologically from the odd-even mass differences involving
the experimental binding energies \cite{ensdf}. Calculations for GT strengths 
are performed for the equilibrium shape of each nucleus, that is, for the 
configuration that minimizes the energy.

To describe GT transitions, a spin-isospin residual separable interaction 
is added to the Skyrme mean field and treated in a deformed proton-neutron 
QRPA. This interaction contains a $ph$ and a $pp$ part. By using separable 
GT forces, the energy eigenvalue problem reduces to find the roots of an 
algebraic equation. The coupling strengths $\chi ^{ph}_{GT}$ and $
\kappa ^{pp}_{GT}$ are chosen phenomenologically to reproduce GT resonances 
and half-lives in different mass regions.
In previous works \cite{sarri-98,sarri-99,sarri-01-odd,sarri-03} we have 
studied the sensitivity of the GT strength distributions to the various 
ingredients contributing to the deformed QRPA calculations, namely to the 
nucleon-nucleon effective force, to pairing correlations, and to residual 
interactions. We found different sensitivities to them. In this work, all 
of these ingredients have been fixed to the most reasonable choices found 
previously. In particular, to be consistent with previous work in the same
mass region \cite{sarri-13}, we use the coupling strengths $\chi ^{ph}_{GT}=0.10$ 
MeV and $\kappa ^{pp}_{GT} = 0.05$ MeV.
The method has been successfully applied in the past to the study of the decay 
properties in different mass regions from proton-rich 
\cite{sarri-09-11,sarri-05-wp,sarri-09-prc} 
to stable \cite{sarri-13,sarri-03,sarri-15-pb} and to neutron-rich nuclei 
\cite{sarri-10}.

The GT strength for a transition from an initial state $i$ to a
final state $f$ is given by

\begin{equation}
B_{if}(GT^{\pm} )= \frac{1}{2J_i+1} \left( \frac{g_A}{g_V} 
\right)_{\rm eff} ^2 \langle f || \sum_j^A \sigma_j t^{\pm}_j 
|| i \rangle ^2 \, ,
\end{equation}
where $(g_A/g_V)_{\rm eff} = 0.7 (g_A/g_V)_{\rm bare}$  is an effective ratio of 
axial and vector coupling factors that takes into account in an effective 
manner the observed quenching of the GT strength. 

When the parent nucleus has an odd nucleon, the ground state can be expressed as 
a one quasiparticle state in which the odd nucleon occupies the single-particle 
orbit of lowest energy. Quasiparticle excitations correspond to configurations
with the odd nucleon in an excited state. 
In the present study we use the equal filling approximation (EFA),
a prescription widely used in mean-field calculations to treat the
dynamics of odd nuclei preserving time-reversal invariance \cite{rayner2}. 
In this approximation the unpaired nucleon is treated in an equal 
footing with its time-reversed state by sitting half a nucleon in a given
orbital and the other half in the time-reversed partner. This approximations has 
been found \cite{schunck10} to be equivalent to the exact blocking
when the time-odd fields of the energy density functional are neglected 
and then it is sufficiently precise for most practical applications.
As we shall see in the next section, because the spin and parity of
the ground and low-lying excited states of the nuclei studied are known 
experimentally, we have chosen these assignments for the odd nucleons to 
describe the corresponding states. In all cases the observed states have 
corresponding calculated states lying in the vicinity of the Fermi 
energy. 

Then, two types of transitions are 
possible. One type is due to phonon excitations in which the odd nucleon acts 
only as a spectator. These are three-quasiparticle states (3qp). In the intrinsic 
frame, the transition amplitudes are basically the same as in the even-even case, 
but with the blocked spectator excluded from the calculation. 
The other type of transitions are those involving the odd nucleon state. These 
are one-quasiparticle states (1qp), which are treated by taking into account 
phonon correlations in the quasiparticle transitions in first order perturbation. 
The transition amplitudes for the correlated states can be found in 
Refs. \cite{krumlinde-84,muto-89,sarri-01-odd}.

In this work we refer the GT strength distributions to the excitation energy 
in the daughter nucleus. For odd-$A$ nuclei we have to deal with 1qp and 3qp 
transitions. The excitation energies for 1qp transitions in the case of an 
odd-neutron nucleus are

\begin{equation}
E_{\mbox{\scriptsize{ex,1qp}}\, [(Z,N-1)\rightarrow (Z-1,N)]}=E_\pi-E_{\pi_0} \, . 
\label{eex1qp}
\end{equation}
where $E_{\pi_0}$ is the lowest proton quasiparticle energy. The excitation 
energy with respect to the ground state of the daughter nucleus for 3qp 
transitions is given by

\begin{equation}
E_{\mbox{\scriptsize{ex,3qp}}\, [(Z,N-1)\rightarrow (Z-1,N)]}=
\omega +E_{\nu,\mbox{\scriptsize{spect}}}-E_{\pi_0} \, . 
\label{eex3qp}
\end{equation}
where $\omega$ is the QRPA excitation energy  and $E_{\nu,\mbox{\scriptsize{spect}}}$
is the energy of the spectator nucleon. Similar expressions are obtained for 
odd-proton nuclei by changing properly protons into neutrons and vice versa.

\begin{figure}[ht]
\includegraphics[width=0.42\textwidth]{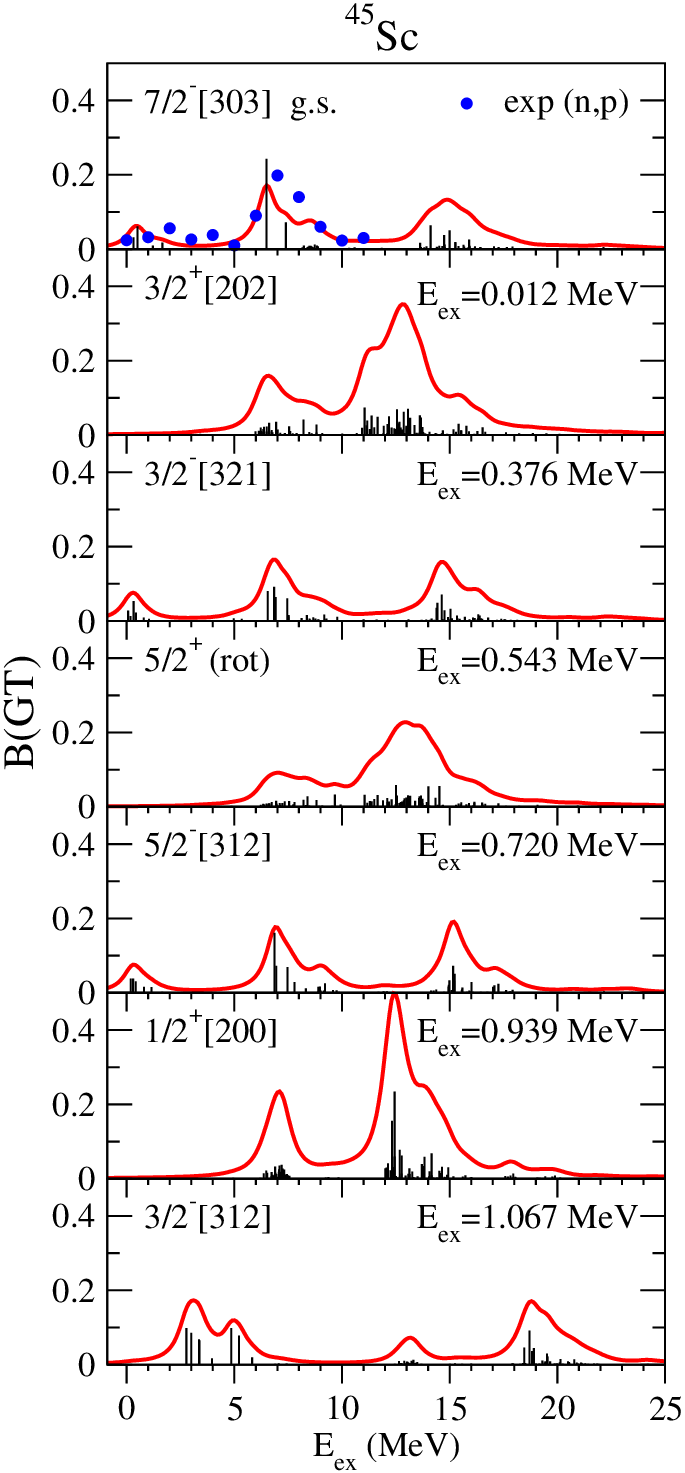}
\caption{(Color online) QRPA Gamow-Teller strength distribution B(GT$^+$) for 
the transition $^{45}$Sc to $^{45}$Ca plotted versus the excitation energy 
of the daughter nucleus. Experimental data in the top panel are compared 
to QRPA results with SLy4 Skyrme force for the decay of the ground state
in $^{45}$Sc. GT strength distributions for the various excited states in
$^{45}$Sc are shown in the lower panels.
Data extracted from $(n,p)$ reactions are from Ref. \cite{alford-91}.} 
\label{bgt_sc45}
\end{figure}

\begin{figure}[ht]
\includegraphics[width=0.42\textwidth]{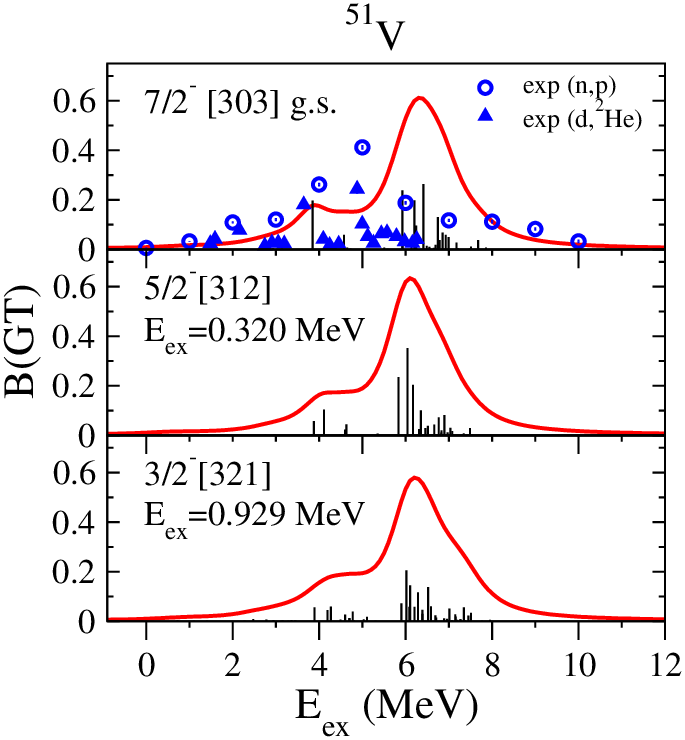}
\caption{(Color online) Similar to Fig. \ref{bgt_sc45} for 
the transition $^{51}$V to $^{51}$Ti. 
Data are from $(n,p)$ \cite{alford-93} 
and ($d,^2$He) \cite{baumer-03} reactions.}
\label{bgt_v51}
\end{figure}

\begin{figure}[ht]
\includegraphics[width=0.42\textwidth]{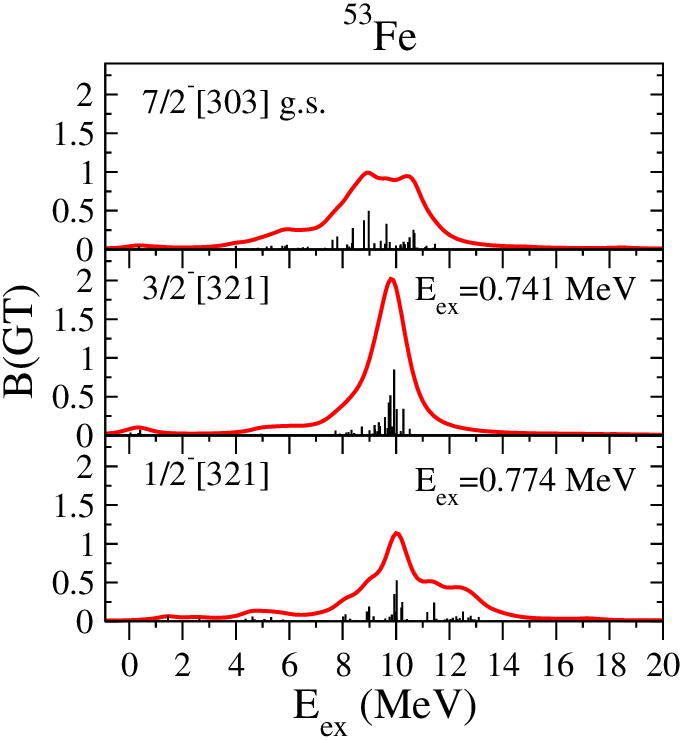}
\caption{(Color online) Similar to Fig. \ref{bgt_sc45} for 
the transition $^{53}$Fe to $^{53}$Mn.}
\label{bgt_fe53}
\end{figure}

\begin{figure}[ht]
\includegraphics[width=0.42\textwidth]{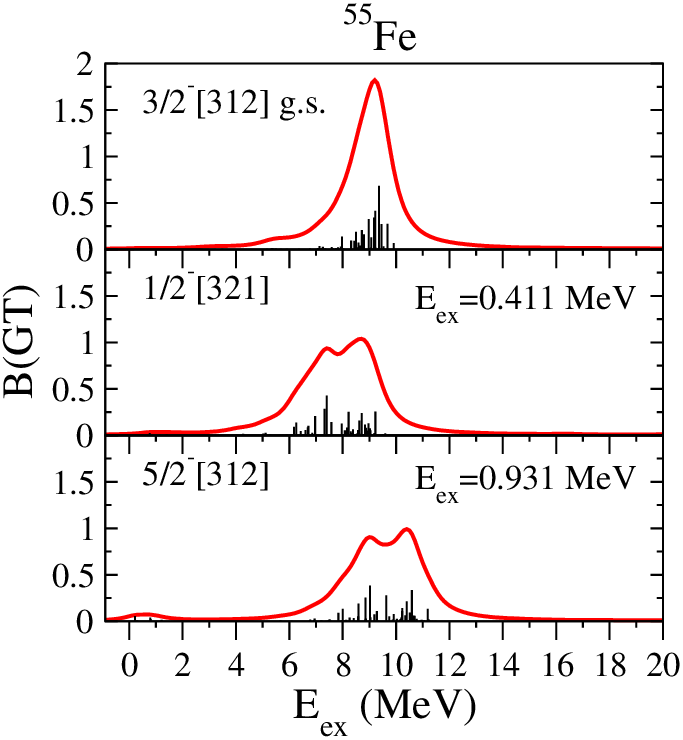}
\caption{(Color online) Similar to Fig. \ref{bgt_sc45} for 
the transition $^{55}$Fe to $^{55}$Mn.}
\label{bgt_fe55}
\end{figure}

\begin{figure}[ht]
\includegraphics[width=0.42\textwidth]{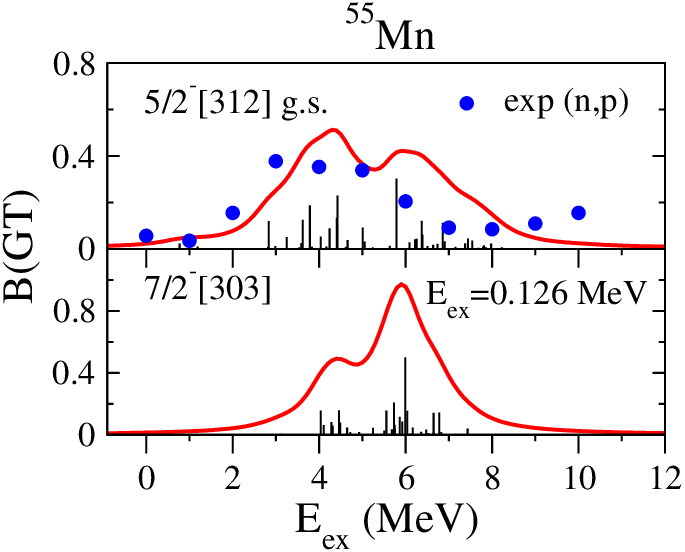}
\caption{(Color online) Similar to Fig. \ref{bgt_sc45} for 
the transition $^{55}$Mn to $^{55}$Cr.
Data extracted from $(n,p)$ reactions are from Ref. \cite{elkateb-94}.} 
\label{bgt_mn55}
\end{figure}

\begin{figure}[ht]
\includegraphics[width=0.42\textwidth]{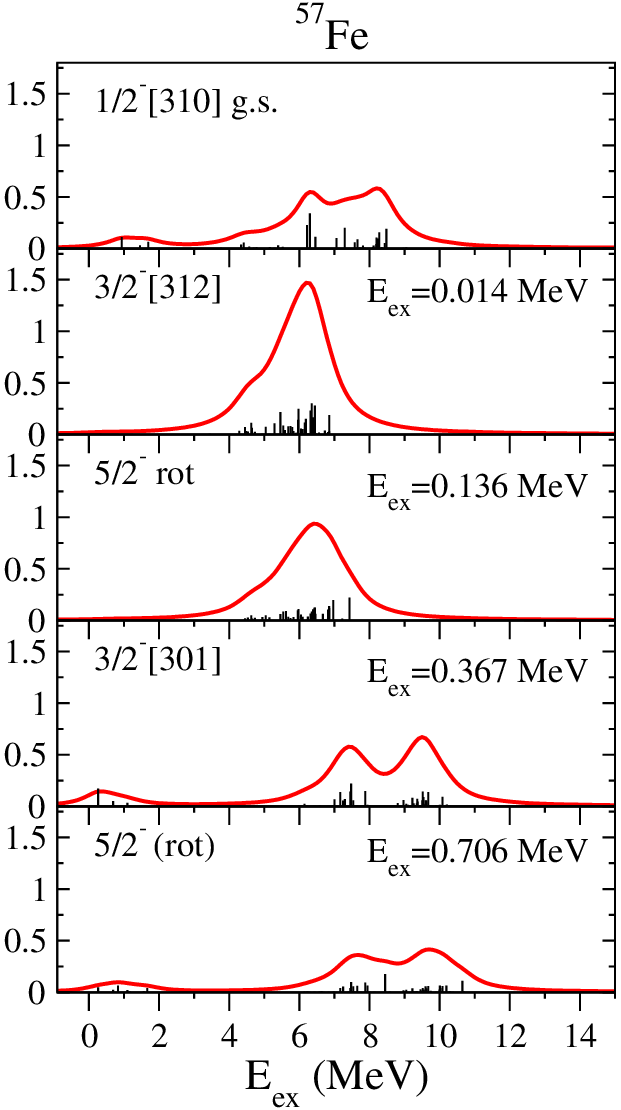}
\caption{(Color online) Similar to Fig. \ref{bgt_sc45} for 
the transition $^{57}$Fe to $^{57}$Mn.}
\label{bgt_fe57}
\end{figure}

\begin{figure}[ht]
\includegraphics[width=0.45\textwidth]{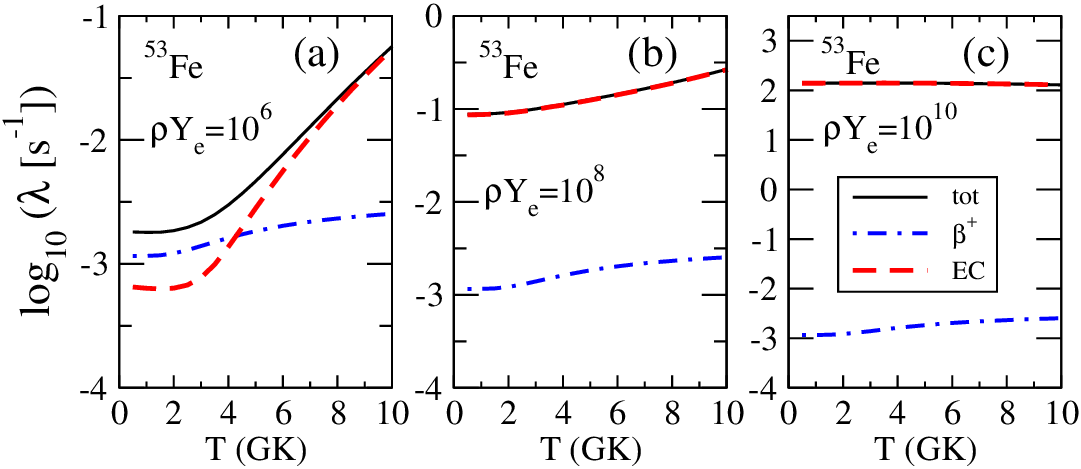}
\caption{(Color online) Weak-decay rates in $^{53}$Fe as a function of T, 
decomposed into their EC and $\beta^+$ contributions for $\rho Y_e = 10^6$ (a),
$\rho Y_e = 10^8$ (b), and $\rho Y_e = 10^{10}$ mol/cm$^3$ (c).} 
\label{beta_ec_fe53}
\end{figure}

\begin{figure}[ht]
\includegraphics[width=0.42\textwidth]{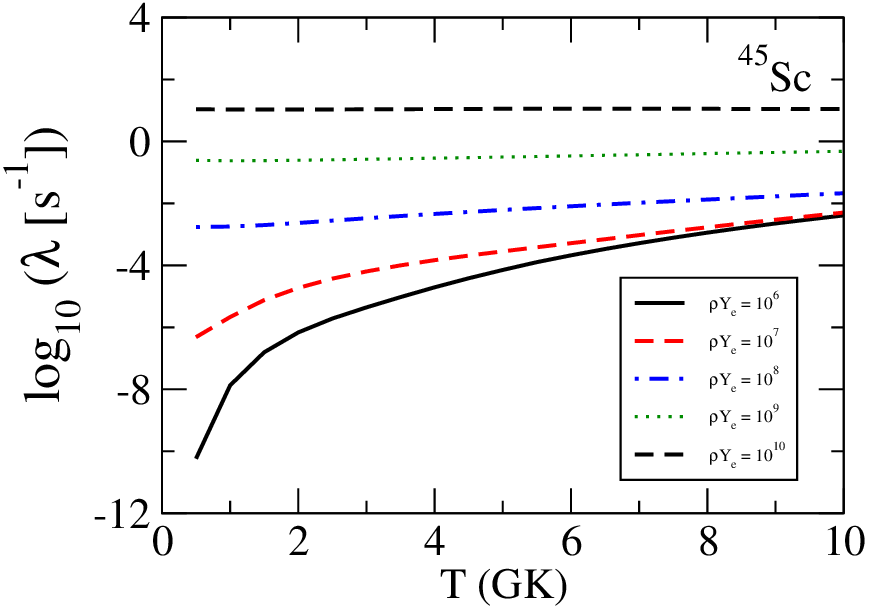}
\caption{(Color online) Total weak-interaction rates for $^{45}$Sc
as a function of T for various densities: $\rho Y_e = 10^6$ (solid black),
$\rho Y_e = 10^7$ (dashed red), $\rho Y_e = 10^8$ (dash-dotted blue), 
$\rho Y_e = 10^9$ (dotted green), and  $\rho Y_e = 10^{10}$ (dashed black).} 
\label{dens_sc45}
\end{figure}

\begin{figure}[ht]
\includegraphics[width=0.42\textwidth]{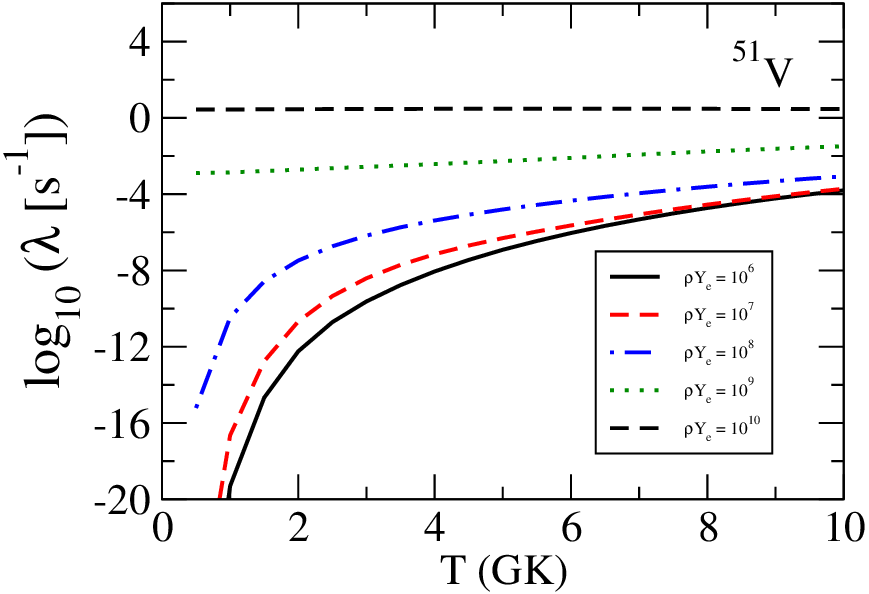}
\caption{(Color online) Same as in Fig. \ref{dens_sc45}, but for $^{51}$V.}  
\label{dens_v51}
\end{figure}

\begin{figure}[ht]
\includegraphics[width=0.42\textwidth]{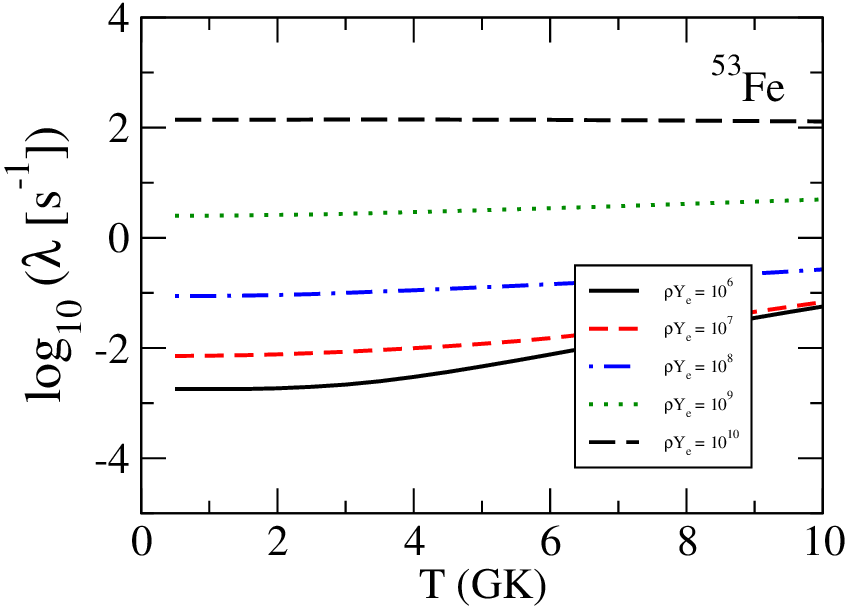}
\caption{(Color online) Same as in Fig. \ref{dens_sc45}, but for $^{53}$Fe.} 
\label{dens_fe53}
\end{figure}

\clearpage

\begin{figure}[ht]
\includegraphics[width=0.42\textwidth]{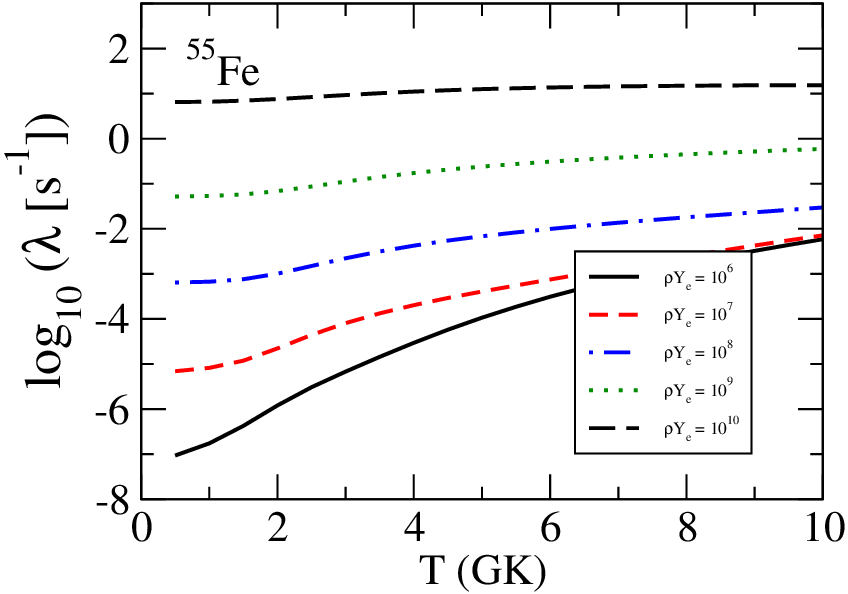}
\caption{(Color online) Same as in Fig. \ref{dens_sc45}, but for $^{55}$Fe.} 
\label{dens_fe55}
\end{figure}

\begin{figure}[ht]
\includegraphics[width=0.42\textwidth]{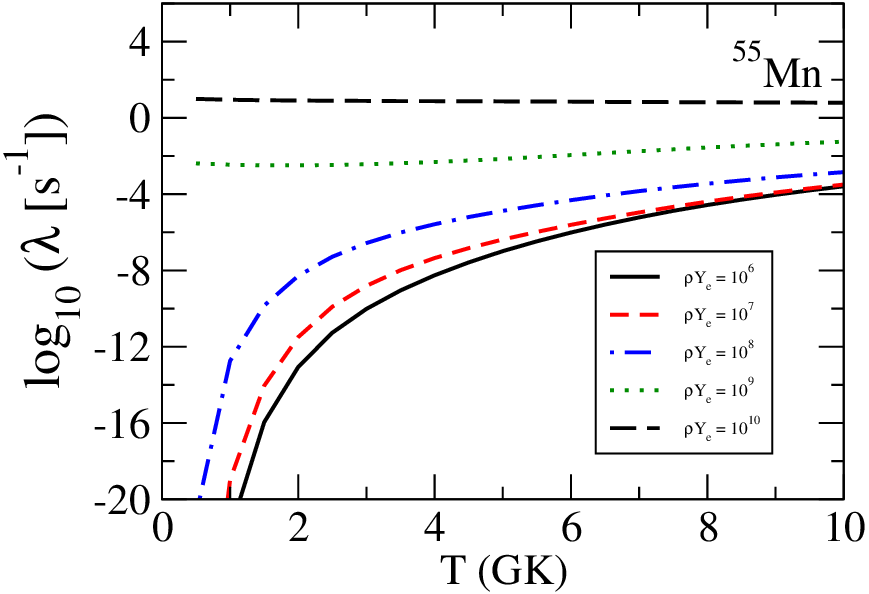}
\caption{(Color online) Same as in Fig. \ref{dens_sc45}, but for $^{55}$Mn.} 
\label{dens_mn55}
\end{figure}

\begin{figure}[ht]
\includegraphics[width=0.42\textwidth]{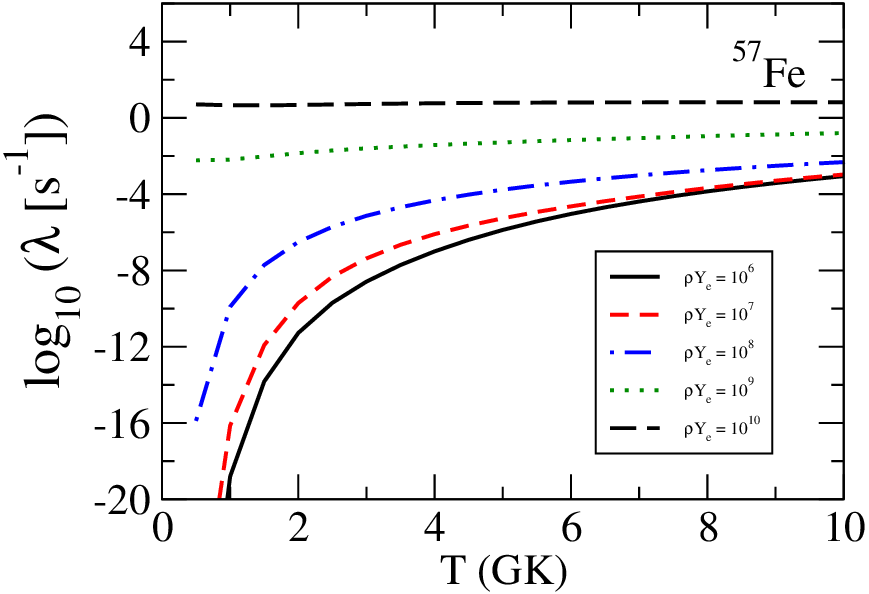}
\caption{(Color online) Same as in Fig. \ref{dens_sc45}, but for $^{57}$Fe.} 
\label{dens_fe57}
\end{figure}

\begin{figure}[ht]
\includegraphics[width=0.42\textwidth]{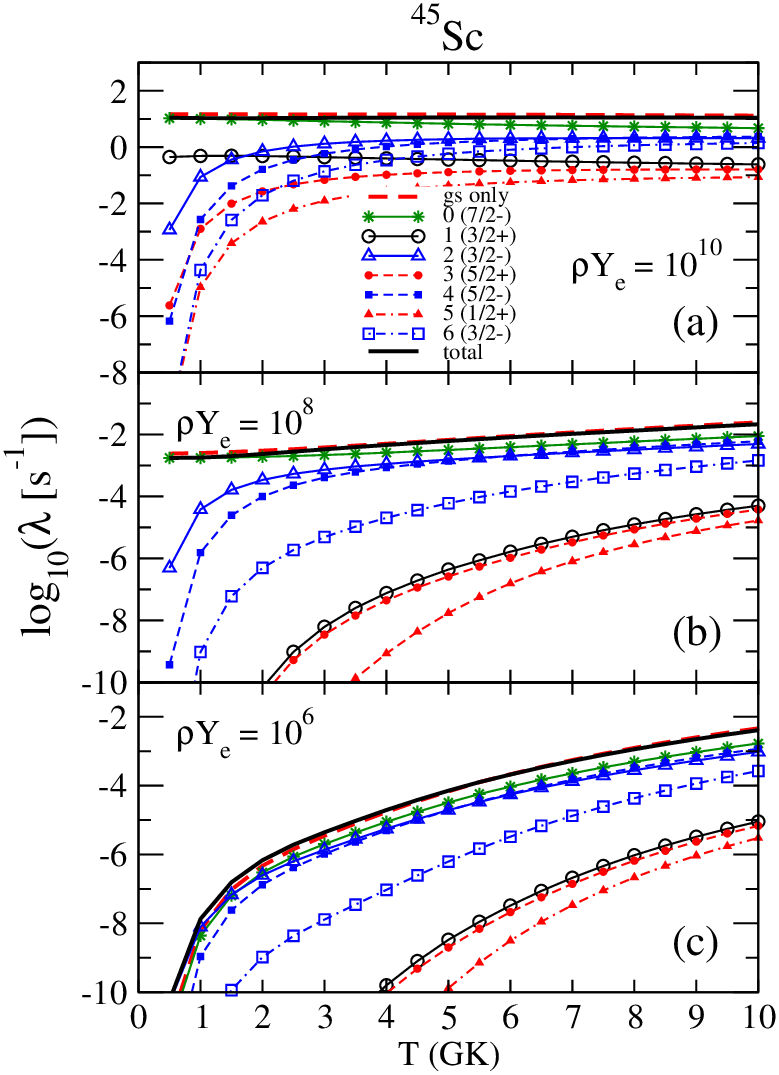}
\caption{(Color online) Weak-decay rates for $^{45}$Sc from SLy4-QRPA
calculations as a function of T (GK) for densities 
$\rho Y_e = 10^{10},10^{8},10^{6}$ mol/cm$^{3}$ in sub-figures (a), (b), and
(c), respectively. The separate contributions from the ground state and 
the various excited states below 1 MeV are also shown.} 
\label{rate_sc45}
\end{figure}

\begin{figure}[ht]
\includegraphics[width=0.42\textwidth]{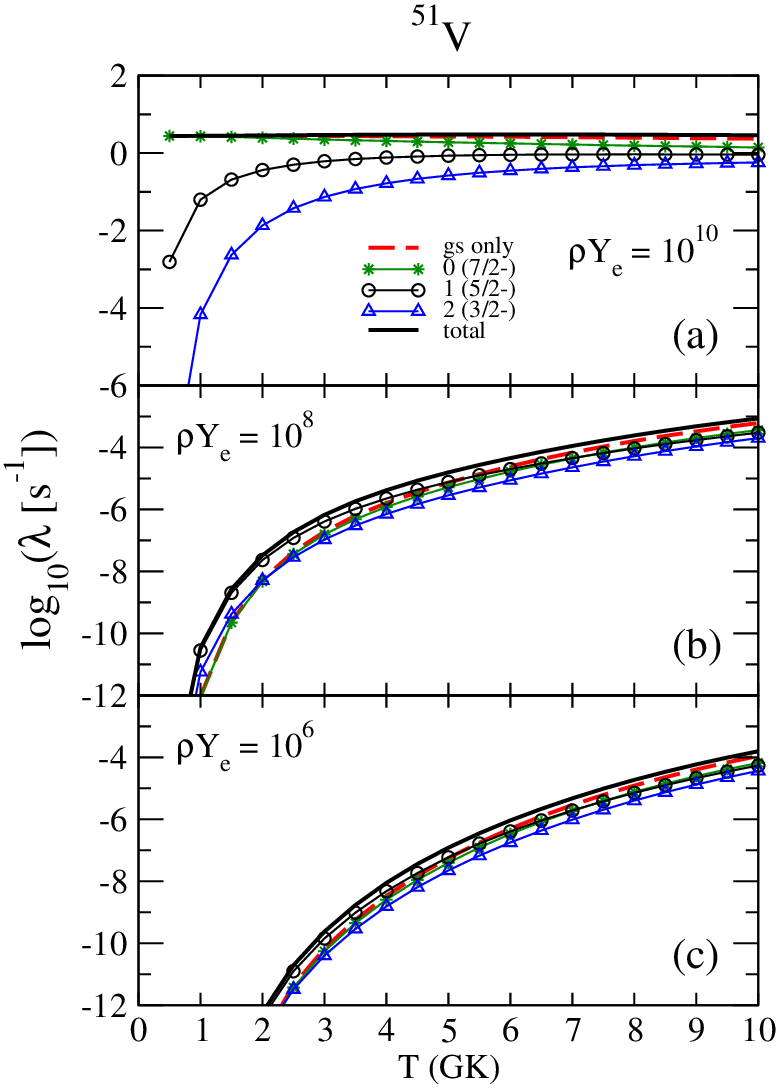}
\caption{(Color online) Same as in Fig. \ref{rate_sc45}, but for $^{51}$V.} 
\label{rate_v51}
\end{figure}

\begin{figure}[ht]
\includegraphics[width=0.42\textwidth]{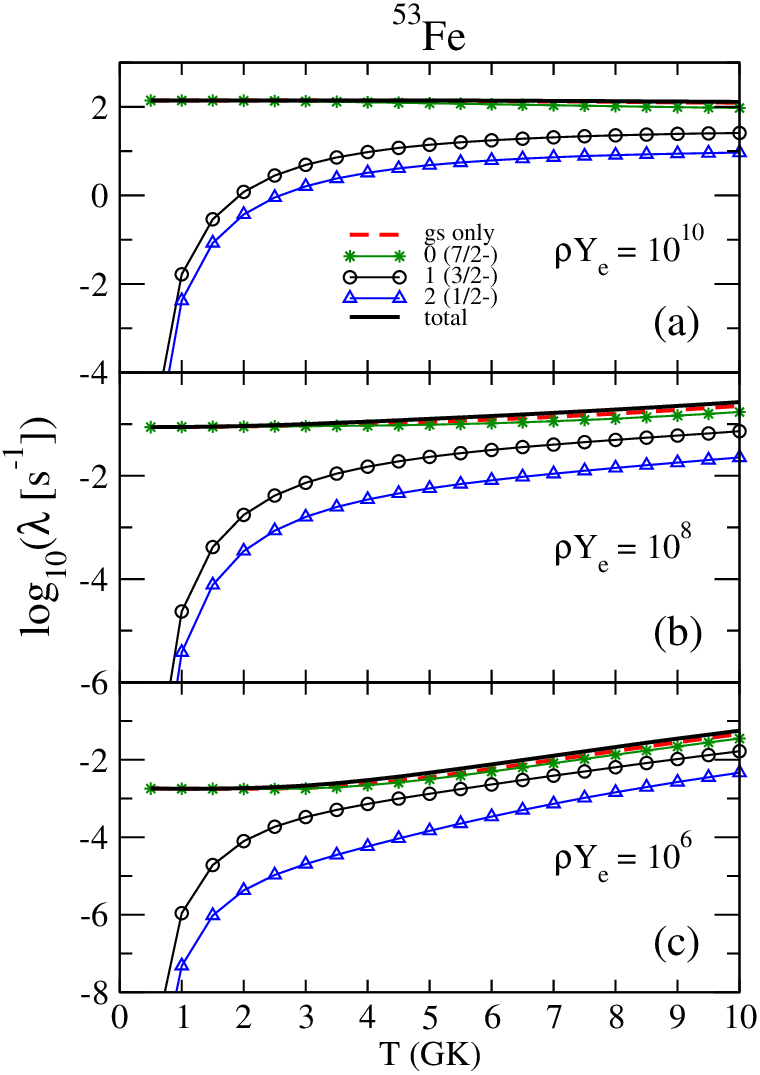}
\caption{(Color online) Same as in Fig. \ref{rate_sc45}, but for $^{53}$Fe.} 
\label{rate_fe53}
\end{figure}

\begin{figure}[ht]
\includegraphics[width=0.42\textwidth]{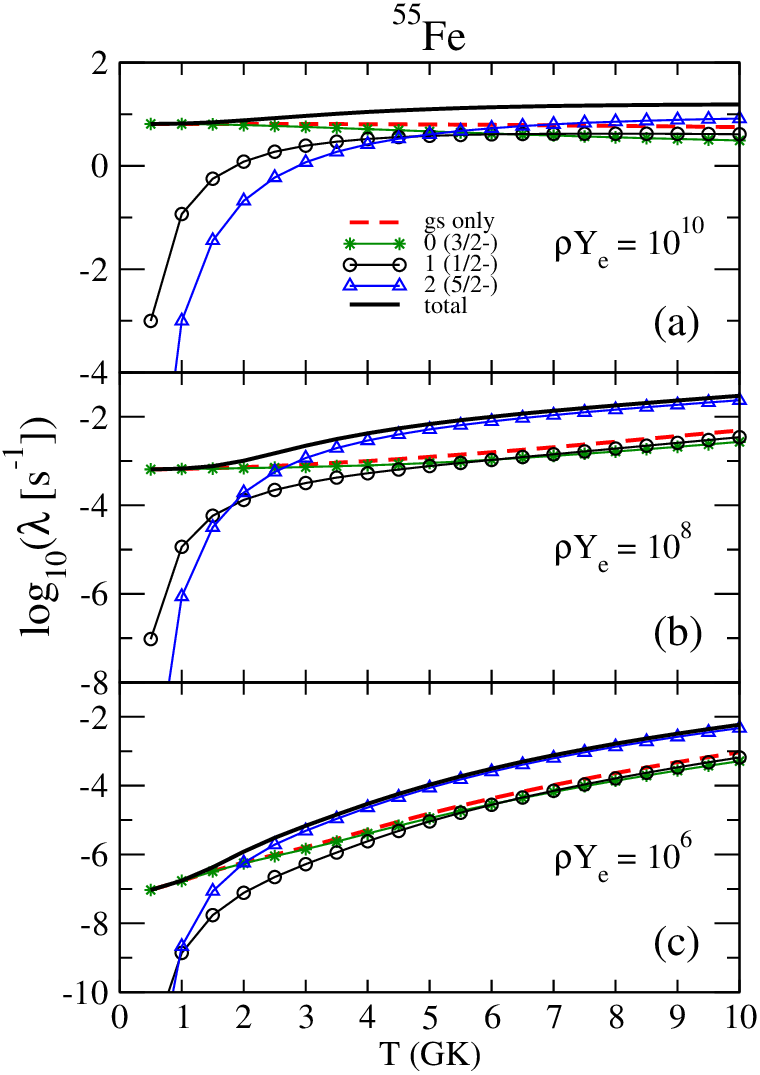}
\caption{(Color online) Same as in Fig. \ref{rate_sc45}, but for $^{55}$Fe.} 
\label{rate_fe55}
\end{figure}

\begin{figure}[ht]
\includegraphics[width=0.42\textwidth]{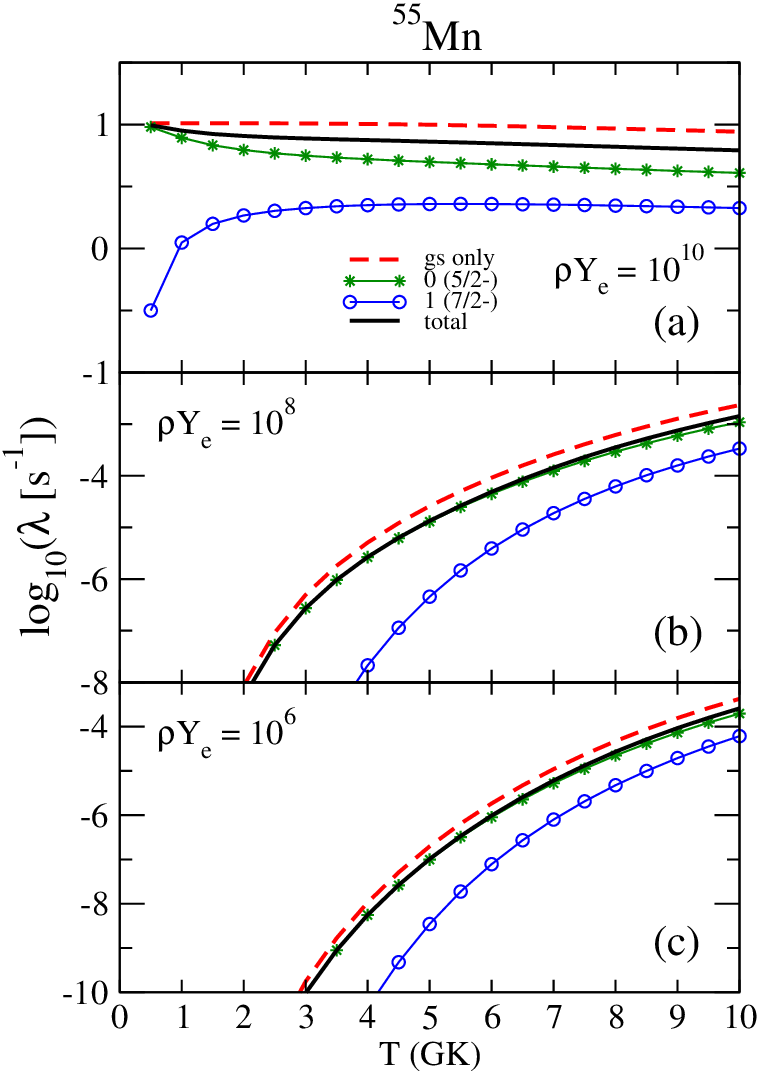}
\caption{(Color online) Same as in Fig. \ref{rate_sc45}, but for $^{55}$Mn.} 
\label{rate_mn55}
\end{figure}

\begin{figure}[ht]
\includegraphics[width=0.42\textwidth]{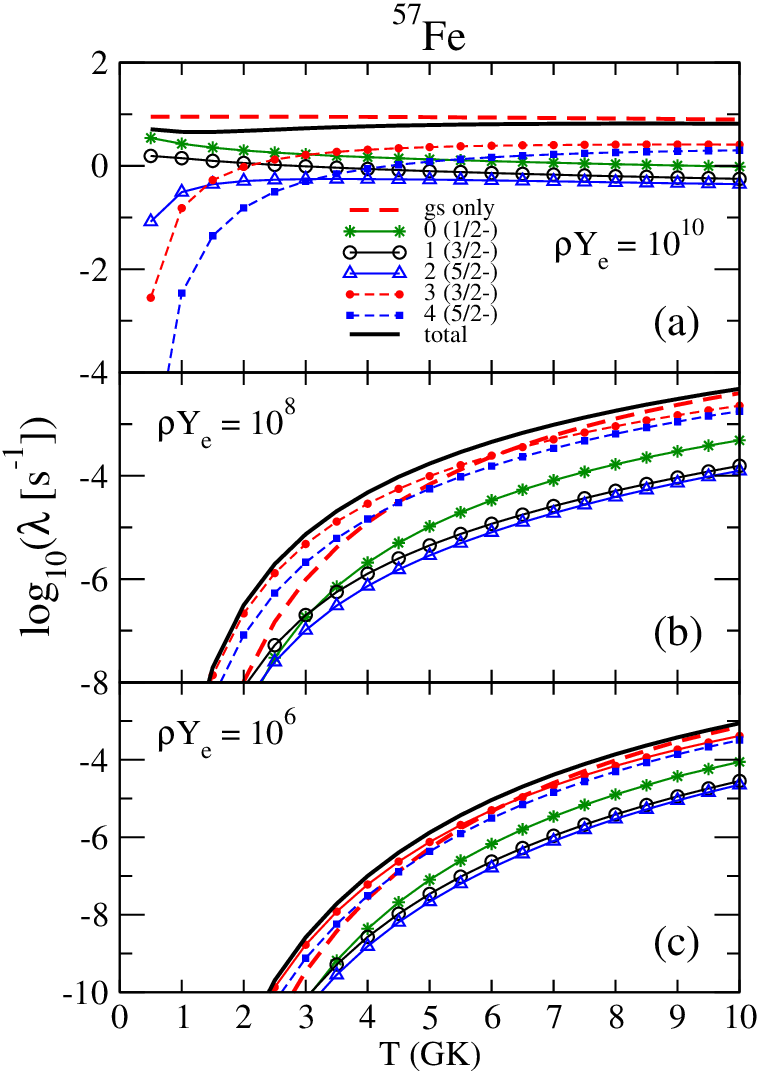}
\caption{(Color online)  Same as in Fig. \ref{rate_sc45}, but for $^{57}$Fe.} 
\label{rate_fe57}
\end{figure}


\section{Results}
\label{results}

\subsection{Gamow-Teller strength distributions}

This section starts with the study of the GT strength distributions corresponding
to the ground state and the various excited states considered. The experimental 
energy spectra below 1 MeV of the nuclei under study can be seen in Fig. 
\ref{en_exp}. The spin-parity assignments of these states can also be seen in the 
figure. We associate the observed states with the calculated quasiparticle states 
around the Fermi level having the same spin and parity or in some instances with 
rotational states built on them. The experimental excitation energies are adopted 
to calculate the thermal population of these states.
This procedure is valid for the general purpose of estimating the relative 
contributions from ground and low-lying excited states in the parent nucleus to 
the weak-decay rates in astrophysical scenarios. However, this approach cannot
be implemented in those cases where low-lying excited states are experimentally 
unknown.

Practically spherical solutions are obtained in the cases of
$^{51}$V and $^{55}$Fe. The rest of nuclei appear to be somewhat deformed. The 
selfconsistent quadrupole deformation $\beta_2$ of the ground states obtained in
our calculations are as follows:
$\beta_2=-0.06$ in $^{45}$Sc, $\beta_2=0.16$ in $^{53}$Fe, $\beta_2=0.13$ in 
$^{55}$Mn, and $\beta_2=0.18$ in $^{57}$Fe.

The correspondence made between the experimentally observed states and the 
Nilsson asymptotic quantum numbers $[N,n_z,m_\ell]K^\pi$ is as follows:
The ground state $7/2^-$ in $^{45}$Sc is associated to the $[303]7/2^-$
state. The next excited states, ordered by increasing energies, correspond to  
$[202]3/2^+$,  $[321]3/2^-$, rotational $5/2^+$, $[312]5/2^-$, $[200]1/2^+$, 
and $[312]3/2^-$. The third excited state, $5/2^+$, is associated with the 
rotational state $J=5/2^+$,  $K=3/2^+$ built on top of the band-head $[202]3/2^+$, 
and we deal with it properly in the calculation of the GT strength. In the case 
of $^{51}$V, the ground state corresponds to $[303]7/2^-$, whereas the first 
two excited states are associated with the asymptotic states $[312]5/2^-$ and 
$[321]3/2^-$, respectively. Similarly, for $^{53}$Fe we have $[303]7/2^-$ as 
ground state and $[321]3/2^-$ and $[321]1/2^-$ as the two first excited states. 
$^{55}$Fe has $[312]3/2^-$ as ground state and  $[321]1/2^-$ and $[312]5/2^-$ 
as excited states. In the case of $^{55}$Mn only the ground state $[312]5/2^-$ 
and first excited state $[303]7/2^-$ are considered.
In $^{57}$Fe the ground state is associated to a $[310]1/2^-$ state. The first 
excited state is associated with $[312]3/2^-$ and the next excited state $5/2^-$ 
is considered to be a rotational state. The next two excited states correspond 
to $[301]3/2^-$ and a rotational state $5/2^-$ built on it.

Figs. \ref{bgt_sc45} - \ref{bgt_fe57} contain the GT strength distributions 
B(GT$^+$) for the ground and excited states of each nucleus, displayed versus 
the excitation energy of the daughter nucleus. The calculations correspond to 
QRPA results with SLy4 Skyrme force. Data from different charge-exchange 
reaction experiments \cite{alford-91,alford-93,baumer-03,elkateb-94} are also 
shown when available.

One can observe in Fig. \ref{bgt_sc45} quite similar profiles for the GT strength
distributions of the negative parity states with a small peak at very low 
excitation energy, a stronger peak around 7 MeV, and another peak at higher energies.
Only the last excited state departs from this pattern. On the other hand, the
positive parity states show a rather different profile with a small peak at
around 6-7 MeV and a stronger one centered at about 12-13 MeV. The GT strength 
distribution of the $7/2^-$ ground state is compared to data \cite{alford-91} 
extracted from $(n,p)$ charge exchange reactions accumulated in 1 MeV bins. The 
agreement with experiment is quite reasonable, reproducing the two bumped structure 
observed, as well as the total strength in the measured energy range.
The next three figures (Figs. \ref{bgt_v51}, \ref{bgt_fe53}, \ref{bgt_fe55}) 
contain the GT strength distributions of the ground state and two first excited 
states in $^{51}$V, $^{53}$Fe, and $^{55}$Fe, respectively. The profiles observed 
in each case are very similar with peaks at 6 MeV, 10 MeV, and 9 MeV, respectively. 
It is also remarkable the similarity of the GT strength distributions of the
various excited states. Experimental measurements from $(n,p)$ \cite{alford-93}
and $(d,^2$He) \cite{baumer-03} reactions are also shown in the case $^{51}$V.
The GT strength appears concentrated at around 5 MeV, whereas the calculations
predict the bump at somewhat larger energies.
The total GT strength measured in both experiments is quite similar and agrees
with the calculations.
In the case of $^{55}$Mn (Fig. \ref{bgt_mn55}) the GT profile of the ground state 
shows a broad peak centered a 5 MeV with a two-peaked substructure containing 
somewhat more strength in the lower one. The GT distribution of the excited 
state is close to that one with the strength shifted to the peak at higher energy. 
The GT distribution of the ground state is compared to $(n,p)$ data from Ref.
\cite{elkateb-94}. The measured strength has a broad peak centered at 4 MeV, which 
is at somewhat lower energy than the calculations. The total measured strength is 
in agreement with the calculated one.
The GT strength 
distributions in $^{57}$Fe (Fig. \ref{bgt_fe57}) exhibit different behavior when 
comparing the ground state with the excited states. In the ground state we find 
some strength at 1 MeV and a two peak structure with centers at 6 and 8 MeV. 
The first $3/2^-$ excited state and its corresponding rotational state $5/2^-$ 
show a broad peak at 6 MeV, whereas the second $3/2^-$ excited state and its 
corresponding rotational state $5/2^-$ show a small bump at 1 MeV and a two 
bump structure with peaks at 7 and 10 MeV.

\subsection{Stellar weak-decay rates}

In the following figures we present weak-interaction rates for the selected 
{\it pf}-shell nuclei as a function of the temperature and for various electron 
densities. 
The range of $T$ considered varies from $T_9=1$ up to $T_9=10$, in $T_9$ (GK) 
units, whereas the range in $\rho Y_e$ varies from $\rho Y_e=10^6$ mol/cm$^3$ 
up to  $\rho Y_e=10^{10}$ mol/cm$^3$. This grid of $\rho Y_e$ and $T$ includes those 
ranges relevant for astrophysical scenarios related to the silicon-burning 
stage in a presupernova star \cite{aufderheide-94} ($\rho Y_e=10^{7}$ 
mol/cm$^3$ and  $T_9=3$), as well as scenarios related to pre-collapse of the 
core \cite{hix-03} and thermonuclear runaway type Ia supernova \cite{iwamoto-99} 
($\rho Y_e=10^{9}$ mol/cm$^3$ and  $T_9=10$).

In these ranges of densities and temperatures, weak-interaction rates are fully 
dominated by EC. $\beta^+$-decays will always contribute for sufficiently high
$\rho$ and $T$ because of thermal population of excited states beyond $Q_{EC}$
energies. However, for $\rho$ and T values in this work, positron-decay
contributions can be neglected. The most favored cases for $\beta^+$ contributions
are those with positive $Q_{EC}$ values that correspond to $^{53}$Fe and $^{55}$Fe. 
We can see in Fig. \ref{beta_ec_fe53} the decomposition of the total 
weak-interaction rates in $^{53}$Fe
into their EC and $\beta^+$ components. At $\rho Y_e = 10^6$  
there is a competition between EC and $\beta^+$ at low T. At higher T, EC clearly 
dominates because it increases while $\beta^+$ remains almost constant. At higher 
densities, the electron chemical potential increases and the electron Fermi-Dirac 
distribution (\ref{se}) allows for higher energy electrons with the result
of a remarkably enhanced EC rate. On the contrary, the $\beta^+$ phase-space factor 
in Eq. (\ref{phiec}) remains invariant with the density and so does the corresponding
$\beta^+$ rate. The final result is that $\beta^+$ is always very small in 
comparison with EC and can be safely neglected. The dominant EC determines the 
total rates in practically all the cases. Therefore, only the total rates will 
be shown in the following figures.

Figs. \ref{dens_sc45} - \ref{dens_fe57} show the total rates as a function of the 
temperature for various densities from $\rho Y_e=10^{6}$ up to $\rho Y_e=10^{10}$
mol/cm$^3$. In these figures one can see that the rates always increase when 
either $\rho$ or T increases. This is a simple consequence of the phase factors
that increase accordingly.
The sensitivity of the EC rates to the environmental $(\rho ,T)$ conditions is 
obvious. At low $\rho$ (low Fermi energies) and low $T$ (sharp shape of the energy 
distribution of the electrons $S_e$), the rates are low and very sensitive to 
details of the GT strength of the low-lying excitations and therefore to 
nuclear structure model calculations. On the other hand, when $\rho$ increases
(larger Fermi energies) and $T$ increases (smearing of the electron distribution
functions), EC rates also increase and become sensitive to all the spectrum, in 
particular, to the centroids of the GT strength distribution and to the total strength.
Then, the whole description of the GT strength distribution is more important than 
a detailed description of the low-lying spectrum.
This is also the reason why all the rates become flat and roughly independent of T 
at high $\rho$. The dependence on the $Q_{if}(Q_{EC})$ values is also apparent.
These energies determine the lower integration limit of the EC phase
factor in Eq. (\ref{phiec}) and thus, the minimum energy of electrons in the stellar
electron plasma to suffer a nuclear capture.
At low densities, where this effect is more pronounced, one can roughly distinguish 
two types of patterns. Those that start with very low rates at low T and increase 
rapidly with T, such as the rates in  $^{45}$Sc,  $^{51}$V, $^{55}$Mn, and $^{57}$Fe, 
and those exhibiting a rather flat behavior with much higher rates, as 
in the cases of $^{53}$Fe and $^{55}$Fe. This global behavior can be 
understood in terms of the $Q_{EC}$ energies. Large negative values, as in the 
cases of $^{51}$V, $^{55}$Mn, and $^{57}$Fe, lead to very low rates.
$Q_{EC}$ energies close to zero as in  $^{45}$Sc lead to intermediate 
rates, and positive values of $Q_{EC}$ cause much higher rates.
This general behavior and sensitivity of the rates to both density and temperature
has been observed in different calculations in this mass region from SM
\cite{dean-98,langanke-00,suzuki-09}  to RPA 
\cite{nabi-99,nabi-09,paar-09,fantina-12,dzhioev-10,niu-11} calculations.
A comparison of the rates obtained within this
formalism and other SM and RPA calculations for even-even {\it pf}-shell nuclei 
can be found in Ref. \cite{sarri-13}.

In the next figures, Figs. \ref{rate_sc45} - \ref{rate_fe57}, the relative
contributions of the ground and excited states to the total rates are studied. 
In addition, total rates and rates obtained from the ground state with full 
population labeled 'gs only' in the figures, are included. The latter are always
larger than the contribution from the ground state at a finite temperature 
(label 0 in the figures) because of the depopulation of the ground state and both
of them converge at zero temperature. As T increases, one would expect a departure
of these two curves (0 and gs only) that would be more apparent when very low-lying
excited states are present. This is the case to a more or less extent in 
$^{45}$Sc, $^{55}$Mn, and $^{57}$Fe (see Fig. \ref{en_exp}). The fact that $^{57}$Fe
shows a stronger effect is due to the angular momentum of the ground state. 
Whereas $^{57}$Fe is a 1/2 state, $^{45}$Sc is a 7/2 state, which is more robust
against depopulation.
Obviously the total contribution (total in the figures) is always larger than 
the contribution from the state 0, but can be larger or smaller than 'gs only',
depending on how the various states become populated.
The relative importance of the various excited states is determined by their 
thermal population at a given T, by the phase factor, and by the structure of 
the GT strength distribution.

In the case of $^{45}$Sc six excited states have been considered. At low densities
($\rho Y_e=10^{6} - 10^{8}$ mol/cm$^3$), the electron chemical potential $\mu_e$
is relatively small and the electrons can only excite low energy states in the
daughter nucleus. In this case the contribution from the positive parity states 
labeled 1, 3, and 5, are very small because these states show very little GT 
strength at low energy in Fig. \ref{bgt_sc45}. Larger rates are found for the states 
labeled 0, 2, and 4 that show some GT strength at very low excitation energies.
At higher densities ($\rho Y_e=10^{10}$ mol/cm$^3$), $\mu_e$ is high enough to allow
high energetic electrons that excite most of the GT strength distribution. In this
case the contribution of the various excited states to the total rates are basically
determined by their thermal population, especially at low temperature.

In the case of $^{51}$V in Fig. \ref{rate_v51} two excited states are considered.
The GT strength distributions of the ground and excited states are quite similar
and therefore the rates are mainly determined by the thermal population of the 
states. We find the contributions to the total rate ordered following the excitation
energies of the parent nucleus, especially at high densities.

In Fig. \ref{rate_fe53} for $^{53}$Fe the three states studied present also very similar 
GT distributions. The two excited states appear at a relatively large energy and their 
thermal population is not significant at these temperatures. The rates are ordered as a 
function of the population of the states.
It is also worth mentioning that the ground state contribution does not fall at low T,
even at low densities, because of the large and positive $Q_{EC}$ energies that allow 
EC at small electron energies.
Similar arguments can be applied to the case of $^{55}$Fe with the difference that 
in this case the rates are lower than the rates of $^{53}$Fe mainly because of the 
smaller $Q_{EC}$ values. Small $Q_{EC}$ values make the rates sensitive to the 
structure of the GT at low excitation energy. This is the reason why the second excited 
state, exhibiting some strength at low energy, contributes more significantly to the rate.
In the case of $^{55}$Mn in  Fig. \ref{rate_mn55} the negative value of $Q_{EC}$ makes
the rates small with contributions ordered according to the population. 

Finally, $^{57}$Fe in Fig. \ref{rate_fe57} shows a competition between the thermal 
population and the GT strength distribution of the various states. 
The large negative $Q_{EC}$ value make the rates very small and quite sensitive to the 
low-lying GT strength, especially at low densities and temperature. Thus, we observe 
that the contribution of the states labeled 3 and 4, which have some GT strength
at low energies, are comparatively large in spite
of their lower population.

\section{Conclusions}
\label{conclusions}

In this work we have evaluated weak-interaction rates, which are basically 
determined by continuum electron-capture rates, at different density and 
temperature conditions holding in stellar scenarios. This study is performed 
on a set of odd-$A$ {\it pf}-shell nuclei representative of the constituents in 
presupernova formations (i.e., $^{45}$Sc, $^{51}$V, $^{53}$Fe, $^{55}$Fe, $^{55}$Mn, 
and $^{57}$Fe). The nuclear structure involved in the calculation of the energy 
distribution of the Gamow-Teller strength is described within a selfconsistent 
deformed HF+BCS+QRPA formalism with density-dependent effective Skyrme 
interactions and spin-isospin residual interactions. This formalism has been shown 
to provide a good description of the decay properties of nuclei by comparing GT 
strength distributions and half-lives with the measured quantities in the 
laboratory under terrestrial conditions. Certainly, the calculated rates are 
subject to uncertainties and there is room for improvement in many theoretical 
aspects. However, the relative importance of the various contributions to the 
total rates studied in this paper may still be of great interest.

First, GT strength distributions for ground states and excited states below
1 MeV excitation energies have been studied in those nuclei and compared with
the available data from charge-exchange reactions. Secondly, weak-interaction
rates have been calculated for ground and excited states at densities and 
temperatures holding in astrophysical scenarios of interest, such as silicon 
burning and presupernova stages. 
Although ECs are always dominant, positron decays have been also included because
some nuclei are unstable $Q_\beta > 0$ and because excited states above $Q_\beta$ 
values can be thermally populated.
Excited states in the parent nucleus beyond 1 MeV have not been considered because 
they are not expected to play any role at the temperatures considered.

The contributions of the excited states to the rates depend on the density and 
temperature through the population of these states and their
phase factors, but depend also on their GT structure.
All in all, we find quite similar results for the total rates and for the rates 
considering only the contribution from the ground states fully populated.
The reason is that population of excited states leads to a depopulation of
ground states and both effects are compensated to some extent when the GT
strength distributions are not very different from each other.

As a general conclusion, we can say that the effect on the rates due to the 
involvement of the excited states of the parent nuclei is of similar order or
smaller than the uncertainties inherent to the nuclear structure. That is,
SM versus QRPA calculations, different nuclear interactions, or different 
treatments in QRPA calculations.
Thus, at the densities and temperatures considered, it is relatively safe to use 
only contributions from ground states, even for odd-$A$ nuclei where
quasiparticle states appear at low energy.


\acknowledgments

This work was supported in part by MINECO (Spain) under Research Grant 
FIS2014--51971--P.

\end{document}